# Performance Indicators Contributing To Success At The Group And Play-Off Stages Of The 2019 Rugby World Cup

R. P. Bunker[a*] and K. Spencer[b*]

[a]RIKEN Center for Advanced Intelligence Project, 1-4-1 Nihonbashi, Chuo-ku, Tokyo, Japan 103-0027. ORCID: 0000-0001-9243-7063

[b]Sport Performance Research Institute New Zealand, Auckland University of Technology, Auckland 0632, New Zealand. ORCHID: 0000-0002-3188-2179

## Abstract

Performance indicators that contributed to success at the group and play-off stages of the 2019 Rugby World Cup were analysed using publicly available data obtained from the official tournament website using both a non-parametric statistical technique, Wilcoxon's signed rank test, and a decision rules technique from machine learning called RIPPER. Our statistical results found that ball carry effectiveness (percentage of ball carries that penetrated the opposition gain-line) and total metres gained (kick metres plus carry metres) were found to contribute to success at both stages of the tournament and that indicators that contributed to success during group stages (dominating possession, making more ball carries, making more passes, winning more rucks, and making less tackles) did not contribute to success at the play-off stage. Our results using RIPPER found that low ball carries and a low lineout success percentage jointly contributed to losing at the group stage, while winning a low number of rucks and carrying over the gain-line a sufficient number of times contributed to winning at the play-off stage of the tournament. The results emphasise the need for teams to adapt their playing strategies from group stage to the play-off stage at the tournament in order to be successful.

## Keywords

Decision Rules; Rugby Union; Sport Performance Analysis; Machine Learning; Non-parametric Statistics; RIPPER

# 1. Introduction

The Rugby World Cup (RWC) is a major global sports event that is held every four years and involves the top 20 rugby countries. The RWC was first held in 1987 yet the 2019 tournament was first one to be held in Japan. South Africa captured the title for a third time, equalling New Zealand's record. Matches during the group stage of the 2019 RWC were noticeably closer than at previous tournaments, perhaps reflective of the narrowing performance gap between Tier one and Tier two (higher ranked and lower ranked) nations.

Over the past few decades, rugby has evolved considerably, with World Rugby implementing various law changes with the objective of promoting changes in playing behaviour such as increasing running with the ball in order to increase the attractiveness of the game to spectators. For instance, compared to the 1995 RWC, ball-in-play time at the 2011 RWC increased by 33%, the number of passes made increased by ~50%, the frequency of rucks/mauls more than doubled, the number of kicks was reduced by 50%, and the frequency of scrums per match decreased from 27 to 17 (Vaz, Vasilica, Kraak & Arrones, 2015). This suggests that the rule changes have impacted on performance.

Performance indicators (M.D. Hughes & Bartlett, 2002) have been constructed using data derived from notational analysis and analysed for some time in rugby union, primarily using descriptive statistics as well as formal statistical methods. Early studies on RWC tournaments made use of notational analysis on match videos, were largely descriptive and generally did not make use of formal statistical methods. M. Hughes and White (1997) investigated differences in the patterns of play of the forwards of successful and unsuccessful teams at the 1991 RWC. They found that the forwards of successful teams dominated lineouts by having a greater variety of lineout options, were technically superior in winning more scrums per match, and were dominant at ruck and maul time. Stanhope and M. Hughes (1997) examined team performances from the 1991 RWC, investigating the different ways points were scored in terms of their tactical significance to successful teams. Although similar patterns of play were generally observed between successful and unsuccessful teams, the noticeable differences were that the successful teams were superior at ruck time and in their kicking performance; in particular, successful teams kicked into danger areas of the field from which to launch attacks. The authors also found that dominant rucking and kicking games also produced more penalties for successful teams, which they were able to take advantage of in danger areas of the field. McCorry, Saunders, O'Donoghue, and Murphy (2001) found that the possession gain to loss, that is, turnovers won or lost, reflected the final ranking of the semi-finalists at the 1995 RWC. Hunter and O'Donoghue (2001) assessed positive and negative aspects of attacking and defensive play, changes in possession, and methods used to gain territory at the 1999 RWC. They found that winning and losing sides differed in the frequency with which they penetrated the opposition's last third of the field, and the frequency of attacking plays in which they outflanked the opposition team. In another study using coded match footage from the 48 tournament matches, it was found that ruck frequency was a strong predictor of success at the 2007 RWC (van Rooyen, Diedrick, & Noakes, 2010). This study was largely descriptive and didn't make use of formal statistical methods. Play-off and group stage matches were considered separately, a similar approach to the present study. It was found that 100% of the play-off matches were won by teams with a lower number of rucks; however, higher ruck frequency was associated with success during the group stage matches. This suggested that the tournament format may influence a team's tactics, and the PIs that are important to success may differ at different stages of the tournament. Vaz et al. (2015) used video from the 12 matches involving New Zealand and France at the 2011 RWC, along with data obtained from the official RWC website and rugbystats.com.au. The data was obtained from coded match footage. They found that there were significant differences in PIs for these two teams. New Zealand's points resulted mostly from tries, while France scored their points mostly from penalties, and there were differences in the performance levels between match halves.

Over the past two decades, data from RWC tournaments has increasingly been made available online, and some studies have augmented data from notational analysis with online-sourced data. There has also been increased focus on ensuring the reliability and validity of data obtained via notational analysis, and on making use of formal statistical methods. In a study by van Rooyen, Lambert, & Noakes (2006), PI data was obtained from the International Rugby Board (IRB) official website and these were augmented with other variables related to movements, which were derived from video analysis of match play. The aim of the study was to explain the performance of four teams at the 2003 RWC. The authors compared South Africa, who lost at the quarter final stage, with the top three teams (England, New Zealand and Australia). The dataset that compared the performance of the teams was analysed using the Kruskal-Wallis test. The data related to the field location of movements were analysed for statistical significance using the Chi-squared test, a one-tailed t-test, and Spearman's rank order correlation coefficients. The most important variables found were the amount of time teams were in possession of the ball, the number of points scored in the second half, and the loss of possession in areas of the field in which the opposition team was likely to score from. In a study using video analysis and Chi-squared tests, van Rooyen and Noakes (2006) found that movement time was an important predictor of success at the 2003 RWC. In particular, they found that a team's ability to construct movements that lasted longer than 1 minute and 20 seconds was important in determining where teams finished at the tournament. Bishop and Barnes (2013) investigated PIs that discriminated between winning and losing teams at the play-off stage of the 2011 RWC. Their data was obtained via coded match footage, which was then checked for intra-observer by a second analyst coding the same matches, as well as re-analysis reliability two weeks later. Several of the PIs were non-normally distributed, thus, the non-parametric Wilcoxon signed rank test (Wilcoxon, 1945) was used. Statistically significant differences between two performance indicators were found, with winning teams conceding a higher percentage of their penalties between the 50 metre and the opposition 22 metre line, and winning teams kicking the ball out of hand more than losing teams. Notably, they also found that successful teams played a more territory-based game as opposed to a possession-based approach. A. Hughes, Barnes, Churchill, and Stone (2017) used data derived from coded match footage, which was coded by two analysts for reliability, and compared PIs that discriminated between winning and losing in the play-off stages of the men's 2015 RWC and 2014 women's RWC. The Shapiro-Wilkes test was used to check normality, and it was found that 91% of the variables were normally distributed; therefore, the parametric two-way mixed ANOVA was used to test for differences between winning and losing teams as well as between genders. They found that in the men's 2015 RWC, winning teams kicked a higher percentage of possession in the opposition 22 to 50 metre zone of the field, for the purpose of creating territory-related pressure (women's was found to be more possession oriented), and the percentage of lineouts won on the opposition throw was found to discriminate between winners and losers.

Researchers have stressed the necessity of more advanced analytical methods in performance analysis for rugby union. M. T. Hughes et al. (2012), Watson, Durbach, Hendricks, and Stewart (2017) and Coughlan, Mountifield, Sharpe, and Mara (2019) highlighted that there are limitations to the univariate analysis of frequency-based PIs, and given the complex nature of rugby, there is a need for greater use of advanced analytical methods. Machine learning (ML) is a relatively new field of study that considers advanced analytical methods, and combines various disciplines including artificial intelligence, computer science, data mining and statistics. ML techniques are increasingly being applied in many disciplines including sport; however, their application to performance analysis in rugby union has only recently begun to be investigated. Recently, Bennett, Bezodis, Shearer, and

Kilduff (2020) built random forest classification models (Breiman, 2001) on PIs from the 40 matches at the group stage of the 2015 RWC, and then used these to predict win/loss outcomes for matches at the play-off stage. The authors found that 13 PIs were significant in predicting the outcome of matches at the group stage: tackle-ratio, clean breaks, average carry, lineouts won, penalties conceded, missed tackles, lineouts won in the opposition 22, defenders beaten, metres carried, kicks from hand, lineout success, penalties in opposition 22, and scrums won. Random forest models with a single variable: tackle ratio, clean breaks or average carry, were found to be able to predict 75%, 70% and 73% of matches at the group stage, respectively. A random forest model built on the group stage data could correctly predict seven out of the eight matches at the play-off stage. Clean breaks alone predicted seven out of eight matches correctly, and tackle ratio and average carry as the two PIs in the model could correctly predict six out of the eight matches. In another recent study, Watson, Hendricks, Stewart, and Durbach (2020) used convolutional and recurrent neural networks to predict the outcomes, namely territory gain, retaining possession, scoring a try, and conceding/being awarded a penalty, of sequences of play based on the order of the events and the on-field locations in which they took place.

In this study, we apply and compare a commonly used non-parametric statistical technique called the Wilcoxon signed rank test with an ML model that learns interpretable decision rules called RIPPER (W.W. Cohen, 1995) to PI data derived from group-stage and play-off stage matches at the 2019 RWC using publicly available data sourced from the official RWC website that are augmented with additional PIs that we calculate based on the original set of variables. To our knowledge, the comparison of a ML model with a statistical method is yet to be investigated in performance analysis in rugby. The interpretable nature of the decision rules generated by RIPPER is appealing and is an advantage over more black-box techniques like random forests and neural networks.

While PIs from group-stage matches at the 2015 RWC were used by Bennett et al. (2020) as inputs to construct an ML model that predicts results at the play-off stage of the tournament, our study differs in that (other than the fact we consider the 2019 RWC) we construct an ML model on both sets of tournament matches (group stage and play-off stage) but we do not use our model for prediction but rather to describe which indicators were most important for success at each stage of the tournament, and to investigate differences in PIs between winning and losing teams in matches at each of the two tournament stages. Bishop and Barnes (2013) considered the play-off stage of the 2011 tournament but not the group stage. Like van Rooyen et al. (2010) and Bennett et al. (2020), we consider play-off and group stage matches as subsets of the tournament matches. However, unlike Bennett et al. (2020), we do not use the group stage matches to predict the outcomes of the play-off matches, since important PIs at the group stage and the play-off stage can be rather different.

## 2. Material & Methods

### 2.1. Measures

The data consisting of PIs were retrieved from the official RWC 2019 website (rugbyworldcup.com). The data on the official RWC 2019 website was provided by the sports data company Stats Perform/Opta. A breakdown of the website-collected PI variables by game area is presented in Table 1.

It is often useful to express frequency-based PIs as a ratio or percentage of another PI, which can often aid interpretation in practical settings (M.D. Hughes & Bartlett, 2002). Therefore, we augmented the original set of PIs with additional calculated PIs, which are listed in Table 2. Several the attacking PIs were transformed to be based on the number of ball carries. For instance, carry metres per ball carry was calculated as a measure of carry effectiveness. Carries over the gain-line, another measure of carry effectiveness, was represented as a percentage of ball carries. Similarly, defenders beaten, line breaks and offloads were all divided by ball carries.

Out of the 45 matches in the RWC 2019 tournament, 37 were group stage matches,[1] and eight were play-off games: the final, semi-finals, quarterfinals, and bronze play-off. As mentioned, the separate consideration of the group stage and play-off stage is based on the hypothesis that there is likely to be a difference in the strategies that are effective at different stages of the tournament.

As mentioned, in this study, we take two approaches to analyse important PIs at the group and play-off stages of the tournament. The first, which we refer to as the "statistical approach" applies the Wilcoxon signed rank test. The second, which we refer to as the "ML approach" applies the decision rule algorithm RIPPER. To implement the models, it was necessary to structure the input datasets differently for each of the two approaches. For the statistical approach, the input dataset was structured such that there was one record for each of the 45 matches, and two PI variable columns for each of the 48 PI variables[2] for the winning and losing teams in each match. Thus, the dataset for the statistical approach consisted of 45 rows and a total of 96 columns. Our input dataset for the ML approach contained exactly the same information, but was structured differently such that it contained two records for each of the 45 matches in the tournament (i.e., one record each for the winning and losing teams of each match) and a column for each of the 48 PI variables, plus the won/lost class variable. Thus, the dataset for the ML approach consisted of 90 rows and 49 columns. The descriptive statistics, which are the same for both datasets, are presented along with the statistical approach results in Section 3.

### 2.2. Procedures

#### 2.2.1. Statistical Approach

Since PI variables, particularly those based on frequencies, are often positively skewed, the median as well as the means of each PI are reported, along with their minimum and maximum values, and standard deviations. The descriptive statistics were generated in Microsoft Power BI.

---

[1] Three group-stage matches in the tournament were cancelled due to Typhoon Hagibis

[2] Because of data quality concerns, three variables related to turnovers, turnovers won, turnovers won in opposition half, and turnovers won in own half, were excluded. Turnovers won in opposition half plus turnovers won in own half were not found to add up to turnovers won in all cases. An email was sent to World Rugby regarding this; however, no response was received, thus, these three variables were excluded from the analysis.

**Table 1.** Performance indicators collected from the official RWC 2019 website.

| Game Area | Performance Indicator |
|---|---|
| Attack | Points scored |
| Attack | Territory % last 10 minutes of match |
| Attack | Territory % whole match |
| Attack | Possession % whole match |
| Attack | Possession % first half |
| Attack | Carry meters |
| Attack | Ball carries |
| Attack | Ball carries over gainline |
| Attack | Passes made |
| Attack | Defenders beaten |
| Attack | Line breaks made |
| Attack | Offloads made |
| Breakdown | Mauls won |
| Breakdown | Rucks won |
| Kicking | Kicks from hand |
| Kicking | Kick meters |
| Kicking | Kicks regathered |
| Kicking | Kicks to touch |
| Kicking | Kicks charged down |
| Kicking | Kicks |
| Set piece | Set pieces won |
| Set piece | Scrums |
| Set piece | Scrums won |
| Set piece | Scrum success % |
| Set piece | Lineouts |
| Set piece | Lineouts won |
| Set piece | Lineout success % |
| Set piece | Lineout steals |
| Discipline | Penalties conceded |
| Discipline | Red cards |
| Discipline | Yellow cards |
| Defence | Tackles missed |
| Defence | Tackles missed |
| Defence | Tackle success % |

A Shapiro-Wilk test was performed on each of the PI variables for winning and losing teams using RStudio (Team, 2015), and more than a third of these variables were found to be non-normally distributed. Therefore, with the relatively small sample size and the repeated measures of each teams' PIs, it was decided that the non-parametric Wilcoxon signed rank test would be used to analyse statistically significant differences between winning and losing teams. The Wilcoxon signed rank test is non-parametric in that it does not require the performance indicator variables' distributions to be normal.

The magnitude of difference is described with effect sizes (ESs) using Cohen's d (J. Cohen, 1988). While J. Cohen (1988) originally interpreted d (0.2) = small, d (0.5) = medium, d (.8) = large,

Sawilowsky (2009) provided the following rules of thumb for interpretation: d (0.01) = very small, d (0.2) = small, d (0.5) = medium, d (.8) = large, d (1.2) = very large, and d (2.0) = huge.

**Table 2.** Additional Performance indicators calculated based on the variables collected from the RWC website.

| Game area | Performance indicator |
|---|---|
| Attack | Carry meters per ball carry |
| Attack | % of carries over gainline |
| Attack | Defenders beaten per ball carry |
| Attack | Line breaks per ball carry |
| Attack | Offloads per ball carry |
| Kicking | Average meters per kick made |
| Kicking | Kicks regained per kick made |
| Kicking | Kicks to touch per kick made |
| Kicking | Kicks charged per kick |
| Set piece | Lineout steal % |
| Attack | Pass to ball carry ratio % |
| Attack/kicking | Kick meters plus carry meters |
| Attack/kicking | % of meters that came from ball carries |
| Attack/kicking | % of meters that came from kicks |

*2.2.2. Machine Learning Approach*

We utilised a model called RIPPER (W.W. Cohen, 1995), a decision rule algorithm that provides results in the form of rules that can be readily interpreted. Interpretability is important in sport performance analysis as it enables coaches and athletes to gain insight and identify important PIs in the hope of improving future performance. For this reason, we avoided the use of "black-box" ML algorithms such as artificial neural networks, support vector machines and random forests.

A decision rule is a simple if-then statement, which consists of a condition (antecedent) and a prediction (Molnar, 2019). Interpreting a decision rule is straightforward; to predict a new instance, start at the top and check whether the rule is applicable (i.e., the condition matches); if so, the right hand side of the rule represents the prediction for this instance (Molnar, 2019). The last rule is the default rule, which applies when none of the preceding rules have applied to an instance, thus ensuring that there is always a prediction. An advantage of RIPPER is that variable selection is performed automatically, unlike many ML models that require a priori variable selection before constructing the actual model. The utility of RIPPER for feature selection purposes in the context of match result prediction in Basketball was highlighted by Thabtah, Zhang, and Abdelhamid (2019).

Given the structure of our dataset, our problem can be treated as a classification problem in which we aim to classify teams' matches into two classes (win or loss) based on the entire set of PI variables (i.e., the original website-obtained variables in Table 1, augmented with the additional calculated variables in Table 2) for each of the group and play-off stage match datasets. The WEKA ML workbench version 3.9.3 (Hall et al., 2009) was used to construct the models. In particular, the JRip algorithm (WEKA's implementation of RIPPER) was trained on the group stage and play-off stage datasets, and was initially tuned to classify all matches correctly As shown in Figure 1, the minimum number of instances that pertain to each rule was lowered from 2 to 1, and pruning was disabled

(usePruning=False, minNo=1).[3] Note that in this study, we performed this tuning to classify all matches correctly since the model is not used for prediction and we are therefore not concerned about over-fitting. This differs from Bennett et al. (2020), whose purpose was to construct an ML model on group-stage matches that was not overly complex (over-fit) so that it was able to predict the separate set of play-off stage matches with a reasonable degree of accuracy. The purpose of our RIPPER ML model is instead to analyse separately the important PI variables at each of the two stages of the tournament but does not involve any prediction of match results.

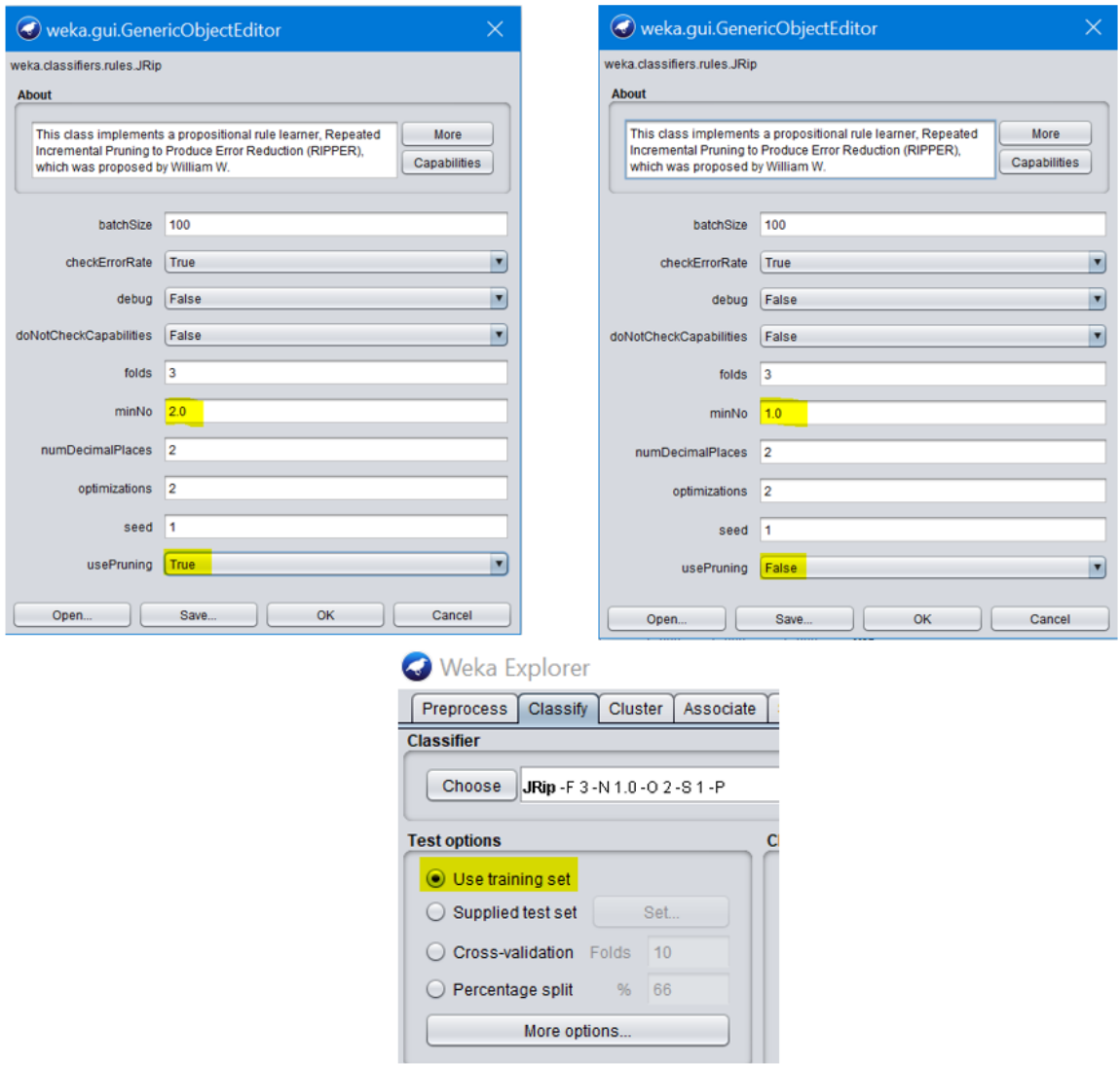

**Figure 1.** JRip (RIPPER) model configuration in WEKA.

---

[3] A J48 decision tree (WEKA's implementation of C4.5 - Quinlan, 1993), which is also readily interpretable, was also trialled, but was found to generate a much larger number of decision rules compared to RIPPER.

# 3. Results

The results for the group stage and play-off matches are presented. The descriptive statistics for each PI, along with the Wilcoxon signed rank test p-values and Cohen's d effect sizes are listed in Tables 3 and 4 for the group stage and play-off stage matches, respectively.

## 3.1. Group Stage Matches

### 3.1.1. Statistical Approach

Differences at the 5% level of significance are discussed in terms of median values. Where there are sets of related variables, e.g., those that relate to ball carries: carries, metres per carry, carries over the gain-line, etc., we identify the variable among this set of PIs with the highest effect size in distinguishing between winning and losing teams.

Winning teams scored 35 points, while losing teams scored 10 points (p = 0.0000). Carry metres had the highest effect size (d = 0.747) of any of the carry-related PI variables. Winning teams in group stage matches dominated carry metres with 537 metres, compared to losing teams who carried 290 metres (p = 0.0000). Clean breaks (in absolute terms) had a higher effect size compared to clean breaks per ball carry. Winning teams broke the opposition line 14 times, more than double that of losing teams who did so 6 times. Defenders beaten (d = 0.660), in absolute terms, had a higher effect size compared to defenders beaten per ball carry. Winning teams beat 32 defenders, double that of losing teams (p = 0.0001). Total metres gained, kick metres plus carry metres, had a statistically significant difference between winning and losing teams, with winning teams gaining 1,224 metres versus 935 metres for losing teams (p = 0.0000). Kicks regathered (d = 0.678), in absolute terms, had a higher effect size compared to kicks regathered per kick. Winning teams in group stage matches regathered 15 kicks compared to 9 for losing teams (p = 0.0000).

The lineout success percentage (d = 0.424) had a higher effect size compared to the number of lineout steals, perhaps suggesting the importance of also retaining ball on team's own throw. Winning teams had a lineout success percentage of 93.3%, compared to losing teams who had 87.5% success (p = 0.0099). Mauls, a common result of lineouts, particularly where teams initiate a driving maul near the opposition try line, were also significantly different between losing teams. Winning teams won five mauls compared to losing teams who won three. Winning teams also won more rucks per match. The number of offloads, in absolute terms, that teams made had a higher effect size compared to the number of offloads made per ball carry. Winning teams made 10 offloads in a match, double that of losing teams. Winning teams in group stage matches made 152 passes compared to losing teams, who made 106. Winning teams also dominated possession, in both the first half and the entire match. Similarly, winning teams had 57% of territory. Winning teams at the group stages of the tournament had higher success at scrum time, winning all their scrums as opposed to 90% of their scrums in the case of losing teams. Of the tackle-related variables, the number of tackles missed in absolute terms had the highest effect size: winning teams missed 16 tackles, half that of losing teams.

**Table 3.** Descriptive statistics and results for performance indicators variables for winning and losing teams across the group stage matches. Wilcoxon signed rank test p-values and Cohen's d effect sizes are reported. The far-right column indicates the sign of the difference in median values for the given performance indicator, i.e., it indicates whether winning teams had higher, lower, or equal values in the performance indicator compared to losing teams. ***, **, and * indicate statistical significance at the 10%, 5% and 1% level, respectively.

| Performance Indicator | Winning (n = 37) | | | | | Losing (n = 37) | | | | | P-Value | d | |
|---|---|---|---|---|---|---|---|---|---|---|---|---|---|
| | Average | Median | Min | Max | StdDev | Average | Median | Min | Max | StdDev | | | |
| points | 38.811 | 35 | 19 | 71 | 12.657 | 11.189 | 10 | 0 | 27 | 7.454 | 0.0000*** | 0.87 | + |
| kick metres plus carry metres | 1215.595 | 1224 | 826 | 1819 | 234.192 | 926.865 | 935 | 525 | 1403 | 216.831 | 0.0000*** | 0.82 | + |
| carry metres | 543.541 | 537 | 285 | 920 | 144.729 | 299.946 | 290 | 150 | 746 | 113.757 | 0.0000*** | 0.75 | + |
| clean breaks | 14.757 | 14 | 5 | 35 | 6.343 | 6.486 | 6 | 0 | 21 | 4.253 | 0.0000*** | 0.73 | + |
| carry metres % of total metres | 0.452 | 0.431 | 0.273 | 0.69 | 0.109 | 0.332 | 0.314 | 0.161 | 0.767 | 0.116 | 0.0000*** | 0.69 | + |
| clean breaks per ball carry | 0.111 | 0.108 | 0.032 | 0.203 | 0.041 | 0.063 | 0.062 | 0 | 0.134 | 0.033 | 0.0000*** | 0.69 | + |
| kicks regathered | 15.378 | 15 | 6 | 26 | 3.872 | 10.297 | 9 | 3 | 20 | 4.832 | 0.0000*** | 0.68 | + |
| carries over gain-line | 48.838 | 45 | 21 | 77 | 14.766 | 31.459 | 30 | 11 | 63 | 10.824 | 0.0000*** | 0.67 | + |
| defenders beaten | 32.324 | 32 | 7 | 53 | 11.228 | 17.838 | 16 | 3 | 48 | 8.512 | 0.0001*** | 0.66 | + |
| tackles missed | 17.838 | 16 | 3 | 48 | 8.512 | 32.297 | 32 | 7 | 53 | 11.188 | 0.0001*** | 0.66 | - |
| carry metres per carry | 4.112 | 4.261 | 1.952 | 5.788 | 0.851 | 3.046 | 2.743 | 1.948 | 5.167 | 0.806 | 0.0001*** | 0.64 | + |
| carries | 132.811 | 128 | 83 | 190 | 24.765 | 99.865 | 104 | 47 | 175 | 27.476 | 0.0001*** | 0.64 | + |
| tackle success % | 0.865 | 0.87 | 0.77 | 0.97 | 0.046 | 0.809 | 0.81 | 0.7 | 0.95 | 0.06 | 0.0001*** | 0.63 | + |
| passes made | 160.514 | 152 | 95 | 254 | 40.461 | 114.568 | 106 | 40 | 253 | 43.151 | 0.0003*** | 0.60 | + |
| possession first half | 0.568 | 0.55 | 0.37 | 0.79 | 0.095 | 0.432 | 0.45 | 0.21 | 0.63 | 0.095 | 0.0003*** | 0.59 | + |
| defenders beaten per ball carry | 0.244 | 0.229 | 0.064 | 0.411 | 0.077 | 0.181 | 0.17 | 0.035 | 0.364 | 0.07 | 0.0004*** | 0.59 | + |
| offloads | 9.838 | 10 | 2 | 24 | 4.705 | 5.486 | 5 | 0 | 21 | 3.839 | 0.0004*** | 0.58 | + |
| kicks regathered per kick | 0.54 | 0.5 | 0.261 | 1.071 | 0.167 | 0.379 | 0.333 | 0.103 | 1.429 | 0.224 | 0.0006*** | 0.56 | + |
| percentage of carries over gain-line | 0.361 | 0.358 | 0.253 | 0.446 | 0.054 | 0.315 | 0.324 | 0.182 | 0.444 | 0.064 | 0.0019*** | 0.51 | + |
| territory | 0.573 | 0.57 | 0.37 | 0.81 | 0.122 | 0.427 | 0.43 | 0.19 | 0.63 | 0.122 | 0.0029*** | 0.49 | + |
| scrum success % | 0.936 | 1 | 0 | 1 | 0.175 | 0.866 | 0.9 | 0 | 1 | 0.197 | 0.0070*** | 0.44 | + |
| offloads made per ball carry | 0.073 | 0.073 | 0.016 | 0.136 | 0.03 | 0.052 | 0.047 | 0 | 0.135 | 0.026 | 0.0072*** | 0.44 | + |
| lineout success % | 0.922 | 0.933 | 0.667 | 1 | 0.083 | 0.851 | 0.875 | 0.5 | 1 | 0.118 | 0.0099*** | 0.42 | + |
| mauls won | 5.054 | 5 | 0 | 10 | 2.66 | 3.324 | 3 | 0 | 7 | 2.08 | 0.0107** | 0.42 | + |
| possession | 0.545 | 0.54 | 0.37 | 0.76 | 0.094 | 0.455 | 0.46 | 0.24 | 0.63 | 0.094 | 0.0111** | 0.42 | + |
| tackles made | 114.324 | 114 | 47 | 183 | 33.123 | 138.757 | 129 | 72 | 218 | 34.687 | 0.0133** | 0.41 | - |
| lineout steals | 1.243 | 1 | 0 | 5 | 1.303 | 0.541 | 0 | 0 | 2 | 0.72 | 0.0231** | 0.37 | + |
| rucks won | 85.541 | 80 | 46 | 127 | 19.699 | 72.405 | 73 | 27 | 122 | 22.436 | 0.0381** | 0.34 | + |
| penalties conceded | 7.514 | 8 | 3 | 15 | 2.637 | 9 | 9 | 3 | 17 | 3.425 | 0.0516 | 0.32 | - |
| scrums | 6.189 | 6 | 0 | 13 | 2.749 | 7.595 | 7 | 0 | 13 | 3.192 | 0.0623* | 0.31 | - |
| pass to ball carry ratio | 1.205 | 1.234 | 0.779 | 1.458 | 0.165 | 1.129 | 1.098 | 0.743 | 1.5 | 0.194 | 0.0726 | 0.3 | + |
| red cards | 0.027 | 0 | 0 | 1 | 0.162 | 0.162 | 0 | 0 | 1 | 0.369 | 0.0726 | 0.3 | = |
| kicks charged per kick | 0.016 | 0 | 0 | 0.065 | 0.02 | 0.028 | 0 | 0 | 0.167 | 0.04 | 0.1134 | 0.26 | = |
| lineout steal % | 0.099 | 0.071 | 0 | 0.5 | 0.105 | 0.066 | 0 | 0 | 0.5 | 0.107 | 0.1287 | 0.25 | + |
| kick metres | 672.054 | 690 | 267 | 1028 | 194.13 | 626.919 | 621 | 223 | 1073 | 208.293 | 0.1698 | 0.23 | + |
| kick metres % of total metres | 0.548 | 0.569 | 0.31 | 0.727 | 0.109 | 0.668 | 0.686 | 0.233 | 0.839 | 0.116 | 0.1698 | 0.23 | - |
| scrums won | 5.973 | 6 | 0 | 13 | 2.746 | 6.919 | 7 | 0 | 13 | 3.123 | 0.1743 | 0.22 | - |
| kicks from hand | 23.838 | 25 | 9 | 39 | 7.023 | 22.216 | 23 | 9 | 39 | 5.946 | 0.1951 | 0.21 | + |
| kicks | 29.757 | 31 | 12 | 47 | 7.084 | 28.054 | 28 | 14 | 42 | 6.089 | 0.2095 | 0.21 | + |
| lineouts won | 11.892 | 13 | 4 | 19 | 3.812 | 10.703 | 11 | 4 | 15 | 3.118 | 0.2934 | 0.17 | + |
| yellow cards | 0.27 | 0 | 0 | 2 | 0.643 | 0.405 | 0 | 0 | 2 | 0.676 | 0.3977 | 0.14 | = |
| kicks charged | 0.514 | 0 | 0 | 2 | 0.683 | 0.703 | 0 | 0 | 3 | 0.926 | 0.4142 | 0.13 | = |
| set pieces won | 18.108 | 18 | 6 | 26 | 4.572 | 17.568 | 17 | 12 | 25 | 3.538 | 0.5437 | 0.10 | + |
| kicks to touch per kick | 0.389 | 0.409 | 0.194 | 0.696 | 0.119 | 0.399 | 0.385 | 0.154 | 0.64 | 0.113 | 0.5831 | 0.09 | + |
| kick metres per kick | 22.565 | 23.226 | 14.74 | 29.03 | 3.875 | 22.009 | 22.429 | 12.39 | 32.16 | 4.313 | 0.5975 | 0.09 | + |
| territory last 10 minutes | 0.496 | 0.42 | 0.1 | 0.99 | 0.244 | 0.504 | 0.58 | 0.01 | 0.9 | 0.244 | 0.8504 | 0.03 | - |
| lineouts | 12.73 | 13 | 5 | 21 | 3.775 | 12.541 | 12 | 6 | 20 | 3.326 | 0.8775 | 0.03 | + |
| kicks to touch | 11.27 | 11 | 5 | 22 | 3.71 | 11.054 | 11 | 4 | 21 | 3.518 | 0.9122 | 0.02 | = |

### 3.1.2. Machine Learning Approach

According to the decision rules generated by RIPPER (Figure 2), losing teams at the group stage had low carry metres (less than or equal to 343 metres) and a lineout success percentage less than or equal to 93.3%. This conjunctive rule explained 26 out of the 37 (70%) losing teams' matches at the group stage. Teams that had a high number of missed tackles (33 or more) despite having carry metres of 341 or more, also lost, explaining six out of 37 (16%) of the losing teams' matches at the group stage. Four out of the 37 (11%) losing teams' matches were the result of low carry effectiveness and few clean breaks per carry. In Fiji's unexpectedly loss to Uruguay, Fiji had very low kicking metres (227 kicking metres, while the median for losing teams at the group stage was 621 metres).

The rules generated by RIPPER not only automatically identify the important features but also provide an idea of their importance through calculating the percentage of losing (or winning) matches that fall under a given rule. For instance, since 70% of losing teams' matches fell within the first rule generated by RIPPER, we can observe that carry metres and lineout success percentage were, jointly, the most important factors at the group stage of the tournament.

```
JRIP rules:
===========

(carry_metres <= 343) and (lineout_success_% <= 0.933) => result=lost (26.0/0.0)
(tackles_missed >= 33) and (carry_metres >= 341) => result=lost (6.0/0.0)
(clean_breaks_per_ball_carry <= 0.074534) and (percentage_of_carries_over_gainline <= 0.335404) => result=lost (4.0/0.0)
(kick_metres <= 227) => result=lost (1.0/0.0)
 => result=won (37.0/0.0)
```

**Figure 2.** RIPPER-generated rules – group stage matches.

### 3.2. Play-off Stages

*3.2.1. Statistical Approach*

The differences in PIs between winning and losing teams in play-off matches at the 5% and 10% levels of significance are described as are non-statistically significant differences with effect sizes exceeding 0.5

Winning teams (mdn = 29 points) in play-off matches typically scored nearly double the points of losing teams (mdn = 15 points) (p =0.014). On attack, winning teams (mdn = 96.5) made significantly fewer ball carries per match than losing teams (mdn = 131.5) (p = 0.04). Winning teams (mdn = 108.5) also made significantly fewer passes per match than losing teams (mdn = 154.5) (p =0.04), which resulted in no significant difference in the pass to ball carry ratio between winning and losing teams. On defence, interestingly, winning teams (mdn = 150) made significantly more tackles per match than losing teams (mdn = 105) (p = 0.04).

Winning teams had higher total metres from kick metres plus carry metres (mdn = 1126.5 metres) compared to losing teams (mdn = 987.5 metres) (p = 0.08). Winning teams (mdn = 45%) had lower possession (mdn = 55%) than losing teams (p = 0.09). In addition, winning teams won fewer rucks (mdn = 71.5) compared to losing teams (mdn = 96) (p = 0.08).[4] Winning teams (mdn = 35.2%) had a greater percentage of their carries over the gain-line compared to losing teams (mdn = 29.6%) (p = 0.08). In terms of the kicking area of the game, winning teams kicked more from hand (mdn = 29.5 kicks) compared to losing teams (mdn = 22 kicks) (p = 0.05). In addition, winning teams did not have any of their kicks charged, while losing teams had (a median of) one kick charged (p = 0.05); 3.9% of losing teams kicks were charged down (p = 0.06).

Winning teams had superior ball carry effectiveness (mdn = 3.6 metres per carry), gaining more metres per ball carry than losing teams (mdn = 2.2 metres per carry) (p = 0.14, d = 0.52). Winning teams (mdn = 784 metres) also had higher kicking metres (mdn = 565.5 metres) (p = 0.14, d = 0.52). On defence, winning teams (mdn = 90%) had a slightly higher tackle success percentage than losing teams (mdn = 86.5%) (p = 0.12, d = 0.55).

---

[4] Unfortunately, the RWC official website did not contain data related to ruck success percentages, dominant tackles, or statistics by field zone, e.g., 22-halfway etc, which would have allowed for a more in-depth analysis.

**Table 4.** Descriptive statistics and results for performance indicators variables for winning and losing teams across the play-off matches. Wilcoxon signed rank test p-values and Cohen's d effect sizes are reported. The far-right column indicates the sign of the difference in median values for the given performance indicator, i.e., it indicates whether winning teams had higher, lower, or equal values in the performance indicator compared to losing teams. ***, **, and * indicate statistical significance at the 10%, 5% and 1% level, respectively.

| Performance Indicator | Winning (n = 8) | | | | | Losing (n = 8) | | | | | p-value | d | |
|---|---|---|---|---|---|---|---|---|---|---|---|---|---|
| | Average | Median | Min | Max | StdDev | Average | Median | Min | Max | StdDev | | | |
| Points | 30.25 | 29 | 19 | 46 | 10.109 | 13 | 15 | 3 | 19 | 5.099 | 0.0141** | 0.87 | + |
| Carries | 106.125 | 96.5 | 71 | 156 | 30.498 | 138.875 | 131.5 | 114 | 181 | 20.763 | 0.0423** | 0.72 | - |
| passes made | 122.25 | 108.5 | 67 | 185 | 41.4 | 161.75 | 154.5 | 115 | 211 | 36.752 | 0.0423** | 0.72 | - |
| tackles made | 161.5 | 150 | 145 | 206 | 22.344 | 111.75 | 105 | 74 | 164 | 28.195 | 0.0421** | 0.72 | + |
| kicks charged | 0.125 | 0 | 0 | 1 | 0.331 | 1.125 | 1 | 0 | 3 | 1.053 | 0.0545* | 0.68 | - |
| kicks charged per kick | 0.003 | 0 | 0 | 0.025 | 0.008 | 0.034 | 0.039 | 0 | 0.073 | 0.029 | 0.0591* | 0.67 | - |
| kicks from hand | 28 | 29.5 | 18 | 37 | 5.979 | 24.25 | 22 | 15 | 36 | 7.241 | 0.0680* | 0.65 | + |
| kick metres plus carry metres | 1155.25 | 1126.5 | 920 | 1515 | 193.937 | 1021.875 | 987.5 | 682 | 1661 | 307.073 | 0.0801* | 0.62 | + |
| percentage of carries over gain-line | 0.354 | 0.352 | 0.278 | 0.435 | 0.05 | 0.298 | 0.296 | 0.202 | 0.411 | 0.059 | 0.0801* | 0.62 | + |
| rucks won | 74.625 | 71.5 | 48 | 110 | 19.937 | 100.75 | 96 | 87 | 137 | 15.658 | 0.0801* | 0.62 | - |
| possession | 0.448 | 0.45 | 0.36 | 0.56 | 0.062 | 0.553 | 0.55 | 0.44 | 0.64 | 0.062 | 0.0898* | 0.6 | - |
| tackle success % | 0.889 | 0.9 | 0.81 | 0.93 | 0.038 | 0.849 | 0.865 | 0.77 | 0.89 | 0.039 | 0.1226 | 0.55 | + |
| carry metres per carry | 3.683 | 3.576 | 2.762 | 4.754 | 0.641 | 2.623 | 2.2 | 1.418 | 4.149 | 1.014 | 0.1415 | 0.52 | + |
| kick metres | 773.625 | 784 | 505 | 1182 | 203.208 | 647.5 | 565.5 | 391 | 1163 | 249.557 | 0.1415 | 0.52 | + |
| kicks | 32 | 33.5 | 23 | 40 | 5.831 | 29.125 | 27 | 17 | 41 | 8.146 | 0.1422 | 0.52 | + |
| yellow cards | 0.375 | 0 | 0 | 1 | 0.484 | 0 | 0 | 0 | 0 | 0 | 0.1489 | 0.51 | + |
| clean breaks per ball carry | 0.095 | 0.095 | 0.041 | 0.172 | 0.039 | 0.059 | 0.058 | 0.016 | 0.119 | 0.034 | 0.1834 | 0.47 | + |
| defenders beaten per ball carry | 0.187 | 0.175 | 0.135 | 0.279 | 0.045 | 0.144 | 0.135 | 0.096 | 0.221 | 0.041 | 0.1834 | 0.47 | + |
| possession first half | 0.436 | 0.415 | 0.32 | 0.62 | 0.083 | 0.564 | 0.585 | 0.38 | 0.68 | 0.083 | 0.2033 | 0.45 | - |
| kicks to touch per kick | 0.29 | 0.302 | 0.217 | 0.346 | 0.043 | 0.369 | 0.378 | 0.195 | 0.529 | 0.119 | 0.262 | 0.4 | - |
| offloads | 6.125 | 4.5 | 2 | 14 | 4.484 | 9.75 | 11.5 | 2 | 15 | 4.63 | 0.271 | 0.39 | - |
| offloads made per ball carry | 0.052 | 0.042 | 0.022 | 0.098 | 0.028 | 0.069 | 0.08 | 0.018 | 0.107 | 0.032 | 0.2936 | 0.37 | - |
| lineout success % | 0.87 | 0.882 | 0.615 | 1 | 0.113 | 0.941 | 0.967 | 0.818 | 1 | 0.066 | 0.3096 | 0.36 | - |
| territory | 0.464 | 0.44 | 0.38 | 0.62 | 0.083 | 0.536 | 0.56 | 0.38 | 0.62 | 0.083 | 0.3456 | 0.33 | - |
| kick metres per kick | 24.248 | 22.697 | 18.38 | 31.95 | 4.847 | 22.063 | 22.777 | 15.69 | 28.37 | 3.976 | 0.3627 | 0.32 | + |
| mauls won | 4.5 | 3.5 | 1 | 10 | 2.784 | 2.875 | 1.5 | 0 | 7 | 2.803 | 0.4017 | 0.3 | + |
| carries over gain-line | 38.25 | 33.5 | 23 | 64 | 14.411 | 41.75 | 42.5 | 23 | 58 | 11.155 | 0.4461 | 0.27 | - |
| kicks to touch | 9.375 | 9.5 | 5 | 12 | 2.446 | 10.125 | 11 | 5 | 14 | 2.571 | 0.479 | 0.25 | - |
| scrums won | 6.375 | 6 | 3 | 11 | 2.395 | 4.875 | 4 | 3 | 8 | 1.763 | 0.5513 | 0.21 | + |
| kicks regathered per kick | 0.519 | 0.466 | 0.189 | 1 | 0.223 | 0.581 | 0.538 | 0.242 | 1.095 | 0.251 | 0.6241 | 0.17 | - |
| lineouts | 10.375 | 9.5 | 6 | 20 | 4.27 | 11.5 | 12 | 7 | 15 | 2.828 | 0.6215 | 0.17 | - |
| territory last 10 minutes | 0.539 | 0.525 | 0.22 | 0.92 | 0.21 | 0.46 | 0.47 | 0.08 | 0.78 | 0.21 | 0.6241 | 0.17 | + |
| scrums | 6.625 | 6.5 | 3 | 11 | 2.497 | 5.125 | 4 | 3 | 9 | 1.9 | 0.5513 | 0.1 | + |
| set pieces won | 15.875 | 16 | 13 | 21 | 2.368 | 15.125 | 15.5 | 10 | 22 | 3.822 | 0.7998 | 0.09 | + |
| carry metres % of total metres | 0.336 | 0.319 | 0.22 | 0.535 | 0.093 | 0.366 | 0.355 | 0.164 | 0.596 | 0.132 | 0.8336 | 0.07 | - |
| clean breaks | 10.125 | 7.5 | 5 | 21 | 5.644 | 8.375 | 7 | 2 | 16 | 4.872 | 0.8334 | 0.07 | + |
| kick metres % of total metres | 0.664 | 0.681 | 0.465 | 0.78 | 0.093 | 0.634 | 0.645 | 0.404 | 0.836 | 0.132 | 0.8334 | 0.07 | + |
| lineout steal % | 0.074 | 0 | 0 | 0.308 | 0.109 | 0.072 | 0.056 | 0 | 0.2 | 0.076 | 0.8339 | 0.07 | - |
| pass to ball carry ratio | 1.138 | 1.168 | 0.944 | 1.268 | 0.104 | 1.163 | 1.109 | 0.888 | 1.581 | 0.203 | 0.8336 | 0.07 | + |
| defenders beaten | 19.875 | 16.5 | 12 | 34 | 7.928 | 20.375 | 20.5 | 11 | 34 | 7.193 | 0.8885 | 0.05 | - |
| kicks regathered | 16.25 | 15 | 7 | 26 | 6.629 | 15.75 | 14.5 | 8 | 24 | 5.309 | 0.8884 | 0.05 | + |
| tackles missed | 20.375 | 20.5 | 11 | 34 | 7.193 | 19.875 | 16.5 | 12 | 34 | 7.928 | 0.8885 | 0.05 | + |
| penalties conceded | 8.5 | 8 | 6 | 13 | 2.121 | 8.5 | 8 | 5 | 12 | 2.236 | 0.9322 | 0.03 | - |
| carry metres | 381.625 | 356.5 | 275 | 580 | 99.896 | 374.375 | 343 | 173 | 639 | 171.081 | 0.9442 | 0.02 | + |
| lineouts won | 9.5 | 9 | 5 | 18 | 3.841 | 10.25 | 9 | 7 | 15 | 3.031 | 0.9438 | 0.02 | = |
| lineout steals | 0.875 | 0 | 0 | 4 | 1.364 | 0.625 | 0.5 | 0 | 2 | 0.696 | 1 | - | - |
| red cards | 0 | 0 | 0 | 0 | 0 | 0.125 | 0 | 0 | 1 | 0.331 | 1 | - | = |
| scrum success % | 0.968 | 1 | 0.857 | 1 | 0.056 | 0.955 | 1 | 0.75 | 1 | 0.086 | 1 | - | = |

### 3.2.2. Machine Learning Approach

When applied to the play-off matches, the RIPPER-generated rules (Figure 3) showed that winning teams won 78 rucks or less. This was the most important factor at this stage of the tournament, accounting for six out of eight (75%) of the winning teams' matches at the play-off stage. On the other hand, despite winning more than 78 rucks, in two play-off matches (New Zealand in their quarterfinal win against Ireland and England in their semi-final win against New Zealand) the winning teams made 55 or more carries over the gain-line. The territory in the last 10 minutes of the match variable appears to be somewhat less relevant, and indeed, when we manually removed this variable from the set of PIs, only the Wales-New Zealand bronze-play-off match was misclassified (Figure 4).

```
JRIP rules:
===========

(rucks_won <= 78) => result=won (6.0/0.0)
(carries_over_gainline >= 55) and (territory_last_10_mins <= 0.39) => result=won (2.0/0.0)
 => result=lost (8.0/0.0)

Number of Rules : 3
```

**Figure 3.** RIPPER-generated rules – play-off stage matches.

```
JRIP rules:
===========
(rucks_won <= 78) => result=won (6.0/0.0)
(carries_over_gainline >= 55) => result=won (2.0/1.0)
 => result=lost (7.0/0.0)

Number of Rules : 3
```

**Figure 4.** RIPPER-generated rules for the play off matches when the component of the rule related to the territory in the last 10 minutes (territory_last_10_mins <= 0.39) is removed, which is seemingly irrelevant. In this case, Wales losing to New Zealand is incorrectly classified as a win by the "(carries_over_gainline >= 55) => result=won" rule (In this match, the bronze play-off, Wales won 137 rucks and had 58 carries over the gain-line, but lost the match).

# 4. Discussion

As expected, there were differences in the performance indicators that contributed to success at the play-off stage compared to the group stage, which suggests the need for teams to adjust their playing strategies at the play-off stage in order to be successful.

At both the group and play-off stages of the tournament, effective ball carries, as measured by the percentage of ball carries that penetrated the opposition gain-line, as well as total metres gained (kick metres plus carry metres), were found to contribute to success. On the other hand, while dominating possession, carrying the ball more frequently, making more passes, winning more rucks, and making less tackles contributed to success at the group stage of the tournament, the opposite was the case at the play-off stage.

At the group stage of the tournament, in comparing our statistical results to those of Bennett et al. (2020), who studied the 2015 RWC, tackle success, clean breaks, average carry (metres per carry), missed tackles, defenders beaten, carry meters, lineout success were found to be important PIs at the group stages of the 2015 as well as the 2019 tournament. While Bennett et al. (2020) found that lineouts won, penalties conceded, kicks from hand and scrums won were also important at the group stage at the group stage of the 2015 RWC, we found in our statistical results that these PIs were not important in distinguishing successful and unsuccessful teams at the group stage of the 2019 RWC (kicks regathered and scrum success were found to be important at the group stage of the 2019 RWC, however).[5]

At the play-off stage of the tournament, the results of the statistical approach found that winning teams made less carries, made less passes, and won less rucks compared to losing teams. Bennett et al. (2020) found that tackle success, clean breaks, average carry in metres, missed tackles, defenders beaten, carry metres, lineout success were important PIs at the group stage of the 2015 RWC, results that are consistent with those of the present study. We also found that there were no statistically significant differences in the ball-to-carry ratios or total carry meters between winning and losing teams at the play-off stage. Despite being in possession of the ball a lower percentage of the time, winning teams in play-off matches were more effective in their carries in terms of both the percentage of their carries that penetrated the gain-line, as well as in metres gained per ball carry. Winning teams made more kicks out of hand and gained more metres via this tactic, resulting in higher total metres gained through either kicks or ball carries. Winning teams also pressured and occasionally charged down opposition kicks, while losing teams were unable to do so. Interestingly, winning teams made more tackles per match, suggesting spending time on defence was not necessarily detrimental, provided their defence was solid, with winning teams having a slightly higher tackle success percentage. Bishop and Barnes (2013), in studying the play-off stages of the 2011 RWC, found that winning teams played more of a territory and kicking style of game rather than a possession-based game. Although our results did not show a significant difference in territory between winning and losing teams, they do suggest that the 2019 RWC was similar in that a possession-based/pick-and-go type of strategy was not effective at the play-off stage of the tournament. Unlike A. Hughes et al. (2017) who studied the 2015 RWC, we did not find that the percentage of opposition lineouts stolen discriminated between winning and losing teams at the play-off stage of the 2019 tournament.

The most important factors identified at the play-off stage by the ML approach were rucks won and number of carries over the gain-line, which were both also identified as important through the statistical approach. Carry metres being an important factor at the group stage is consistent with the results obtained via the statistical approach, with carry metres found to have the largest effect size (apart from points). However, the decision rules approach also identified a joint relationship in which low carry metres together in conjunction with a low lineout success percentage was the most important factor contributing to losing at the group stage of the tournament, explaining 70% of losing teams' matches at the group stage (lineout success did not have an overly large effect size in the results from the Wilcoxon signed rank test, although it was still significant at the 1% level). A high number of tackles missed, making few clean breaks per carry, and having low carry effectiveness also contributed to losses at the group stage. This again suggests that a possession-based game with a repeated pick-and-go type strategy may not have been effective at the play-off stage of the 2019 RWC, which agrees with the results obtained via the statistical approach.

---

[5] Bennett et al. (2020) also found that penalties and lineouts won in the opposition 22m zone were important at the group stage of the 2015 RWC, however these PIs were unfortunately not available on the 2019 RWC official website.

The results of the ML approach again highlight that forming a large number of rucks was not an advantage at the play-off stage of the tournament, but the two teams that did have a large number could otherwise win through carry effectiveness in terms of having a high number of carries that penetrated the gain-line.

## 5. Conclusion

Our statistical and ML approaches provided somewhat different results, most notably in the number of important PIs identified, with RIPPER selecting a small subset of the original PIs, perhaps a disadvantage since it allows for a less in-depth analysis. The obvious advantages of decision rules are that, compared to statistical approach, which required the calculation of many Wilcoxon signed rank tests for each of the PI variables, the decision rules are fast and easy to generate and interpret. Like non-parametric statistical tests like the Wilcoxon signed rank test, decision rules do not require distributional assumptions such as normality. On the other hand, a weakness of decision rules is that, particularly for a small number of matches (e.g., we only have eight play-off matches), some rules generated may be relatively random in nature, but happen to classify the match outcomes correctly (evident in the "territory in the last 10 minutes" variable appearing in the RIPPER-generated rules for the play-off matches).

The present study is not without limitations. The variables included as performance indicators were limited to those that were available on the 2019 RWC official website. Performance indicators such as dominant tackles, and ruck frequency, ruck success percentage or rucks lost were not available. In addition, variables were not available by field position, e.g., 22 metre line to halfway, 22 metre line to try line, etc. This limited the ability to compare with the results of some prior studies. Also, only team-level performance indicators were considered in the present study, player-level variables were not.

An interesting avenue for future work could be to augment performance indicator variables with external variables such as venue, weather, referees and so on, and conduct a comparative analysis of their importance. Another would be to experiment with other interpretable machine learning models on other similar datasets that consist of performance indicator variables.

# References

Bennett, M., Bezodis, N. E., Shearer, D. A., & Kilduff, L. P. (2020). Predicting performance at the group-phase and knockout-phase of the 2015 rugby world cup. *European Journal of Sport Science*, 1–9.

Bishop, L., & Barnes, A. (2013). Performance indicators that discriminate winning and losing in the knockout stages of the 2011 rugby world cup. *International Journal of Performance Analysis in Sport*, *13* (1), 149–159.

Breiman, L. (2001). Random forests. *Machine learning, 45*(1), 5-32.

Cohen, J. (1988). Statistical power analysis for the behavioral sciences, 2nd ed. Hillsdale, NJ: Erlbaum.

Cohen, W. W. (1995). Fast effective rule induction. In *Machine learning proceedings* 1995 (pp. 115-123). Morgan Kaufmann.

Coughlan, M., Mountifield, C., Sharpe, S., & Mara, J. K. (2019). How they scored the tries: applying cluster analysis to identify playing patterns that lead to tries in super rugby. *International Journal of Performance Analysis in Sport*, 1–17.

Hall, M., Frank, E., Holmes, G., Pfahringer, B., Reutemann, P., & Witten, I. H. (2009). The WEKA data mining software: an update. *ACM SIGKDD explorations newsletter, 11*(1), 10-18.

Hughes, A., Barnes, A., Churchill, S. M., & Stone, J. A. (2017). Performance indicators that discriminate winning and losing in elite men's and women's rugby union. *International Journal of Performance Analysis in Sport, 17* (4), 534–544.

Hughes, M., & White, P. (1997). An analysis of forward play in the 1991 rugby union world cup for men. *Notational analysis of sport I & II*, 183–191.

Hughes, M. D., & Bartlett, R. M. (2002). The use of performance indicators in performance analysis. *Journal of sports sciences, 20* (10), 739–754.

Hughes, M. T., Hughes, M. D., Williams, J., James, N., Vučković, G., & Locke, D. (2012). Performance indicators in rugby union. *Journal of Human Sport & Exercise*.

Hunter, P., & O'Donoghue, P. (2001). A match analysis of the 1999 rugby union world cup. In Books of abstracts fifth world congress of performance analysis in sports (pp. 85–90).

McCorry, M., Saunders, E., O'Donoghue, P., & Murphy, M. (2001). A match analysis of the knockout stages of the 1995 rugby union world cup. *Notational analysis of sport III*. Cardiff: UWIC, 230–239.

Molnar, C. (2019). Interpretable machine learning. *Lulu. com*.

Sawilowsky, S. S. (2009). New effect size rules of thumb. *Journal of Modern Applied Statistical Methods, 8*(2), 26.

Stanhope, J., & Hughes, M. (1997). An analysis of scoring in the 1991 rugby union world cup for men. *Notational Analysis of Sport I y II*, 167–176.

Team, R. O. (2015). RStudio: integrated development for r. *RStudio, Inc., Boston, MA URL http://www.rstudio.com, 42*, 14.

Thabtah, F., Zhang, L., & Abdelhamid, N. (2019). NBA game result prediction using feature analysis and machine learning. *Annals of Data Science, 6*(1), 103-116.

van Rooyen, K. M., Diedrick, E., & Noakes, D. T. (2010). Ruck frequency as a predictor of success in the 2007 rugby world cup tournament. *International Journal of Performance Analysis in Sport, 10* (1), 33–46.

van Rooyen, K. M., Lambert, I. M., & Noakes, D. T. (2006). A retrospective analysis of the IRB statistics and video analysis of match play to explain the performance of four teams in the 2003 rugby world cup. *International Journal of Performance Analysis in Sport, 6* (1), 57–72.

van Rooyen, K. M., & Noakes, D. T. (2006). Movement time as a predictor of success in the 2003 rugby world cup tournament. *International journal of Performance analysis in Sport, 6* (1), 30–39.

Vaz, L., Vasilica, I., Kraak, W., & Arrones, S. L. (2015). Comparison of scoring profile and game related statistics of the two finalists during the different stages of the 2011 rugby world cup. *International Journal of Performance Analysis in Sport, 15* (3), 967–982.

Watson, N., Durbach, I., Hendricks, S., & Stewart, T. (2017). On the validity of team performance indicators in rugby union. International Journal of Performance Analysis in Sport, 17 (4), 609–621.

Watson, N., Hendricks, S., Stewart, T., & Durbach, I. (2020). Integrating machine learning and decision support in tactical decision-making in rugby union. *Journal of the Operational Research Society*, 1-12.

Wilcoxon, F. (1945). Individual Comparisons by Ranking Methods. *Biometrics Bulletin 1*, 6 (1945).